\documentclass[twocolumn,showpacs,amsmath,amssymb,amsfonts,a4paper,aps,prd]{revtex4}
\usepackage{graphicx,color}

\newcommand{\f}[2]{\frac{#1}{#2}}

\def\be{\begin{equation}}
\def\ee{\end{equation}}
\def\bea{\begin{eqnarray}}
\def\eea{\end{eqnarray}}
\def\bl{\begin{align}}
\def\el{\end{align}}

\def\nn{\nonumber \\}

\begin{document}

\title{Further matters in space-time geometry: $f\left(R,T,R_{\mu\nu}T^{\mu\nu}\right)$ gravity}

\author{Zahra Haghani$^1$}
\email{z_haghani@sbu.ac.ir}
\author{Tiberiu Harko$^2$}
\email{t.harko@ucl.ac.uk}
\author{Francisco S. N. Lobo$^3$}
\email{flobo@cii.fc.ul.pt}
\author{Hamid Reza Sepangi$^1$}
\email{hr-sepangi@sbu.ac.ir}
\author{Shahab Shahidi$^1$}
\email{s_shahidi@sbu.ac.ir}
\affiliation{$^1$Department of Physics, Shahid Beheshti University, G. C.,
Evin,Tehran 19839, Iran}
 \affiliation{$^2$Department of Mathematics, University College London, Gower Street,
London, WC1E 6BT, United Kingdom}
\affiliation{$^3$Centro de Astronomia e Astrof\'{\i}sica da
Universidade de Lisboa, Campo Grande, Edificio C8 1749-016 Lisboa,
Portugal}
\begin{abstract}

We consider a gravitational theory in which matter is non-minimally coupled to geometry, with the effective Lagrangian of the gravitational field being given by an arbitrary function of the Ricci scalar, the trace of the matter energy-momentum tensor, and the contraction of the Ricci tensor with the matter energy-momentum tensor. The field equations of the theory are obtained in the metric formalism, and the equation of motion of a massive test particle is derived. In this type of theory the matter energy-momentum tensor is generally not conserved, and this non-conservation determines the appearance of an extra-force acting on the particles in motion in the gravitational field. It is interesting to note that in the present gravitational theory, the extra-force explicitly depends on the Ricci tensor, which entails a relevant deviation from the geodesic motion, especially for strong gravitational fields, thus rendering the possibility of a spacetime curvature enhancement by the $R_{\mu\nu}T^{\mu\nu}$
coupling. The Newtonian limit of the theory is also considered, and an explicit expression for the  extra-acceleration which depends on the matter density is obtained in the small velocity limit for dust particles.
 We also analyze in detail the so-called Dolgov-Kawasaki instability, and obtain the stability conditions of the theory with respect to local perturbations. A particular class of gravitational field equations  can be obtained by  imposing the conservation of the energy-momentum tensor. We derive the corresponding field equations for the conservative case  by using a Lagrange multiplier method, from a gravitational action that explicitly contains an independent parameter multiplying the divergence of the energy-momentum tensor. The cosmological implications of the theory are investigated in detail for both the conservative and non-conservative cases, and several classes of exact analytical and approximate solutions are obtained.

\end{abstract}

\date{\today}

\maketitle

                              \section{Introduction}\label{sec1}

The recently released Planck satellite data of the 2.7 degree Cosmic Microwave Background (CMB) full sky survey \cite{1,2} have generally confirmed the standard $\Lambda $CDM ($\Lambda $Cold Dark Matter) cosmological paradigm. A major goal of the Planck experiment was to test the $\Lambda $CDM
model to high precision, and identify possible areas of tension. In fact, an interesting  result from the fits of the basic $\Lambda$CDM model to the Planck
power spectra is the lower than expected  value of the Hubble constant $H_0$, $H_0=67.3\pm 1.2$ km/s/ Mpc. The Hubble constant can be tightly constrained by CMB data alone in the $\Lambda $CDM model \cite{2}. The Planck data has also further constrained the parameters of dark energy, a possible cause of the late-time cosmic acceleration. Indeed, a central problem in present day physics  is to elucidate the nature of  dark energy, thought to be driving the accelerated expansion of the Universe. Perhaps the most straightforward explanation for dark energy is the presence of a cosmological constant. An alternative is dynamical dark energy \cite{DE, Copeland:2006wr}, usually assumed to be a very light scalar field, having a canonical kinetic energy term, and being minimally coupled to gravity. The cosmological constant $\Lambda $ has an equation of state $w=p/\rho =-1$, where $p$ and $\rho $ are the effective thermodynamic pressure and energy density associated with $\Lambda $,  while scalar field theories
usually have time varying equations of state with $w\geq -1$ \cite{2}.

The CMB alone does not strongly constrain the dark energy equation of state parameter $w$, due to the
two-dimensional geometric degeneracy present in  dark energy models. However, this degeneracy can be broken by combining the CMB data with lower redshift distance measurements \cite{2}. By combining the Planck data with the measurements of $H_0$ in \cite{Riess}, the authors provide an equation of state parameter of dark energy given by $w=-1.24^{+0.18}_{-0.19}$, which is off  by more than the 2$\sigma $ compared to $w=-1$ \cite{2}. The Planck data combined with the  Baryon Acoustic Oscillations (BAO) data give $w=-1.13^{+0.24}_{-0.25}$ \cite{2}. Therefore, presently there is no convincing observational evidence that could clearly establish the nature of dark energy.  Moreover, the accelerated expansion of the Universe (see \cite{Riess} and references therein), the virial mass discrepancy at the galactic cluster level and  the galaxy rotation curves \cite{Bi87} as well as other cosmological observations suggest that the standard general relativistic gravitational field equations, based on the Einstein-Hilbert
action $S=\int{\left(R/2+L_m\right)\sqrt{-g}d^4x}$, where $R$ is the scalar curvature, and $L_m$ is the matter Lagrangian density, cannot describe the Universe at large scales, beside passing the Solar System tests. From a cosmological viewpoint, this amounts to introducing, by hand, the dark matter and dark energy components in the theory, in addition to ordinary matter and energy.

Another possibility is to modify the basic structure of the Einstein-Hilbert action in the hope that such a modification could naturally explain dark matter and dark energy, without  resorting to some exotic forms of matter. Initially, the  interest in the extension of the Einstein-Hilbert action was focused on the modification of the geometric part of the action. One of the interesting research avenues is the introduction of higher order terms to the gravitational field action through the substitution of the Ricci scalar with a generic function $f(R)$ \cite{carroll,staro}. (see also \cite{rev} for a review.) The most
serious difficulty of  $f(R)$ theories is that in general, these theories seem incapable of passing the
standard Solar System tests \cite{badfR}. However, there exists some theories that can accommodate
this problem \cite{goodfR}. The phase space analysis of the general $f(R)$ theories is considered in \cite{phasefR}. One can also generalize $f(R)$ type gravity theories by including the function $f(R)$ in the bulk action of the brane-world theories \cite{shahab,dynam,zahra}.

A new class of modified theories of gravity was proposed recently, consisting of the superposition of the metric Einstein-Hilbert Lagrangian with a $f(\cal R)$ term constructed \`{a} la Palatini \cite{Harko:2011nh}. The dynamically equivalent scalar-tensor representation of the theory was also formulated, and it was shown that even if the scalar field is very light, the theory passes the Solar System observational constraints. Therefore the theory predicts the existence of a long-range scalar field, modifying the cosmology \cite{Capozziello:2012ny,Capozziello:2013wq}, galactic dynamics \cite{Capozziello:2012qt} and wormhole physics \cite{Capozziello:2012hr}.

Another interesting fact of $f(R)$ gravitational theories is that they are equivalent to  Brans-Dicke theories
with a specific $\omega$ parameter \cite{rev}. This suggests that the theory describes the non-minimal
coupling between matter and geometry in the Einstein frame. It also provides a motivation to consider
non-minimal coupling between matter and geometry in a more general manner at the action level.
In the Einstein-Hilbert action, which has a simple additive structure in terms of $R$ and $L_m$,
geometry and matter appear at two very different conceptual levels, without any interaction
between them. However, the idea that the gravitational action may not be additive in matter
and geometry cannot be rejected {\it a priori}.
One of the first efforts in this direction was made in \cite{goenner} where, based on
very general physical arguments,  a framework was suggested in which terms with
non-minimal coupling  between matter and geometry should be expected in the
action. As a consequence, a general action for the gravitational field would require a general
coupling between the Ricci scalar and the matter Lagrangian.

In this context, a maximal extension of the Einstein-Hilbert Lagrangian was introduced in \cite{fRLm}, where the Lagrangian of the gravitational field was considered to be a general function of $R$ and $L_m$ and therefore this theory came to be known as the $f\left(R,L_m\right)$ gravity theory. In theories with non-minimal geometry-matter coupling there exist an extra force, which arises from the interaction between matter and geometry, as initially suggested in \cite{extraforcefR} and \cite{Harko1}, respectively.  This extra force affects the motion of a test point particle, causing it to undergo a
non-geodesic motion \cite{matterlagrangian}. On the other hand it has been suggested that the extra
force could be ignored if one uses a matter Lagrangian of the form $L_m=p$
instead of $L_m=-\rho$ \cite{soti}. However, in \cite{Harko2} it was shown that when the particle number is conserved, the Lagrangian of a barotropic perfect fluid is $\mathcal{L}_m=-\rho [c^2 +\int P(\rho)/\rho^2 d\rho]$, where $\rho$ is the \textit{rest mass} density and $P(\rho)$ is the pressure.
In turn, the $f\left(R,L_m\right)$ theory was generalized recently by considering a gravitational theory with an action given by an arbitrary function of the Ricci scalar, the matter Lagrangian density, a scalar field and a kinetic term constructed from the gradients of the scalar field, respectively \cite{Harko3}.

Another difficulty of the non-minimal theories is that, in general, the equivalence principle is violated. In fact, it has been shown that the observational data of the Abell Cluster A586 exhibits evidence of the interaction between dark matter and dark energy, and that this interaction implies a violation of the Equivalence Principle \cite{berto}. The mass profile in this particular cluster is approximately spherical, and  it is a relaxed cluster, since it has not undergone any important merging process in the last few Gyrs. For the Abell Cluster A586 the kinetic energy $\rho _K$ and the gravitational potential energy $\rho _W$ can be computed. Then the generalized virial theorem $2\rho _K+\rho _W=\xi \rho _w$, where $\xi $ is a coupling constant, allows to estimate the magnitude of the dark energy-dark matter interaction, as well as the degree of violation of the equivalence principle that should be detectable in large scale cluster surveys \cite{berto}.

In the $f\left(R,L_m\right)$ type theories \cite{fRLm,extraforcefR,Harko1,matterlagrangian,soti,Harko2,berto,Harko3} it is assumed that all the properties of the matter are encoded in the matter Lagrangian $L_m$. An alternative view would be to consider theories in which matter, described by some of its thermodynamic parameters, different from the Lagrangian, couple directly to geometry.
In the standard $\Lambda $CDM model the cosmological constant is
spatially uniform and time independent, as required by the principle of general covariance. Physically, it can  be interpreted as a relativistic
ideal fluid obeying the equation of state $p+\rho =0$. Moreover, this cosmological fluid obeys an equation of continuity that does not depend on the matter energy density. Such a form of dark energy is said to be non-interacting \cite{Pop}. An interaction between ordinary matter and dark energy can be introduced in
the form of a time-dependent cosmological constant. However, to preserve the general covariance of the field equations, a variable cosmological constant must depend only on relativistic invariants. The assumption $\Lambda =\Lambda (R)$ leads to the $f(R)$ class of modified gravity theories. In these models, a Legendre - Helmholtz transformation of the Lagrangian, or a conformal transformation of the metric, transforms
the gravitational field equations of $f(R)$ gravity into the form of the Einstein
equations of general relativity, with an additional scalar field. Another choice, in which the cosmological constant is a function of the trace of the energy-momentum tensor $T$, was proposed in \cite{Pop}. One advantage of the choice of a gravitational Lagrangian of the form $R+2\Lambda (T)$, as compared to  $f(R)$-type gravity theories, is that since we use one and the same metric tensor, the problem  about which frame (Einstein or Jordan) is physical, does not appear \cite{Pop}.

Following the initial work done in \cite{Pop}, the general non-minimal coupling between matter and geometry was considered in the framework of a Lagrangian of the form  $f(R,T)$, consisting of an arbitrary function of the Ricci scalar and the trace of the energy-momentum tensor \cite{fRT}. The gravitational field equations in the metric formalism, as well as the equations of motion for test particles, which follow from the covariant divergence of the stress-energy tensor, were obtained.  The equations of motion of  test particles were also obtained from a variational principle. The motion of massive test particles
is non-geodesic and takes place in the presence of an extra force orthogonal to the four-velocity.

The astrophysical and cosmological implications of the $f(R,T)$ gravity theory have been extensively investigated recently \cite{impl}. A reconstruction of the cosmological models in $f(R,T)$ gravity was performed in \cite{Re}.  The dust fluid reproduces the $\Lambda $CDM cosmology, the phantom-non-phantom era, and the phantom cosmology. The numerical simulation for the Hubble parameter shows good agreement with the BAO observational data for low redshifts $z<2$. The study of the evolution of scalar cosmological perturbations was performed \cite{Gom}, by assuming  a specific model that guarantee the standard continuity equation. The complete set of differential equations for the matter density perturbations was obtained and it was shown that for general $f(R,T)$ Lagrangians the quasi-static approximation leads to very different results as compared to the ones derived in the frame of the  $\Lambda$CDM model. For sub-Hubble modes, the density contrast obeys a second order differential equation, with explicit
wave-number dependence, and subsequent strong divergences on the cosmological evolution of the perturbations. A comparison of these results with the usual quasi-static approximation in general relativity  shows that the density contrast quantities evolve very differently. There is also a difference in the linear regime between these theories. The results obtained in \cite{Gom} for $f(R,T)$ gravity are in  contradiction with the usually assumed behavior of the density contrast, and imposes strong limitations on the viability of the $f(R,T)=f_1(R)+f_2(T)$ type models. The growth of the scalar perturbations in the sub-Hubble limit, for this model, is scale-dependent. However, one should emphasize that the observational data provided by the Planck satellite \cite{Planck} show clear evidence of the scale dependence of the CMB power spectrum. On the other hand it seems that  Lagrangians of the form $f(R,T)$ cannot lead in general to the standard energy-momentum conservation equations \cite{fRT, Gom}. Cosmological solutions of $f(R,T)$ modified theories of gravity of the form $g(R)+h(T)$,  $g(R)h(T)$, and $g(R)(1+h(T))$, respectively, for perfect fluids in spatially FLRW metric  were investigated through phase space analysis in \cite{SB}.  Acceptable cosmological solutions, which contain a  matter dominated era, followed by a late-time accelerated expansion, were found.

However, the $f\left(R,L_m\right)$ or $f(R,T)$ type theories are not the most general Lagrangians describing the
non-minimal coupling between matter and  geometry. For example, one may generalize the above modified theories of gravity by introducing a term $R_{\mu\nu}T^{\mu\nu}$ in the Lagrangian. Indeed, examples of such couplings can be found in the Einstein-Born-Infeld theories \cite{deser} when one expands the square root in the Lagrangian.
An interesting difference in $f(R,T)$ gravity and in an inclusion of the $R_{\mu\nu}T^{\mu\nu}$ term, is that in considering a traceless energy-momentum tensor, i.e., $T=0$, the field equations of $f(R,T)$ gravity reduces to those of $f(R)$ gravity theories. However, considering the presence of the $R_{\mu\nu}T^{\mu\nu}$ coupling term still entails a non-minimal coupling to the electromagnetic field.

It is the purpose of this work to consider an extension of the $f(R,T)$ gravity theory by also taking into account a possible coupling between the energy-momentum tensor of  ordinary matter, $T_{\mu \nu}$, and the Ricci curvature tensor $R_{\mu \nu}$. Therefore we propose to describe the gravitational field by means of a Lagrangian of the form $f\left(R,T,R_{\mu\nu}T^{\mu\nu}\right)$ (a similar approach is carried out in \cite{Odintsovetal}, but in a different setting), where $f$ is an arbitrary function in the arguments $R$, $T$, and $R_{\mu\nu}T^{\mu\nu}$, respectively. We obtain the gravitational field equations for this theory, and formulate them as an effective Einstein field equation. The equation of motion of massive test particles is also obtained from the field equations. In this type of theories the energy-momentum tensor is generally non-conserved. In order to study the Newtonian limit of the theory we derive the equation of motion from a variational principle. An important requirement for any
generalized gravity theory, besides passing the Solar System tests, is its stability. Thus, we analyze in detail the
so-called Dolgov-Kawasaki instability, obtaining the stability conditions for the theory. An interesting question is the possibility of the conservation of the energy-momentum tensor in such theories. We impose the conservation of the energy-momentum tensor by employing a Lagrange multiplier method. The gravitational equations with energy-momentum conservation are derived from an action with the Lagrange multiplier, multiplying the energy-momentum tensor, included. The cosmological implications of the theory are investigated for both the conservative and non-conservative cases, and several classes of analytical and numerical solutions are obtained.

The present paper is organized as follows. The gravitational field equations of  $f\left(R,T,R_{\mu\nu}T^{\mu\nu}\right)$ gravity theory are derived in Section \ref{sec2}, and the equations of motion of massive test particles are obtained in Section \ref{sec3}. The Newtonian limit of the theory is studied in Section \ref{sec4}, where in particular, we obtain the generalized Poisson equation. In Section \ref{sec5}, the Dolgov-Kawasaki instability in the $f\left(R,T,R_{\mu\nu}T^{\mu\nu}\right)$ gravity theory is further investigated. In Section \ref{sec6}, the field equations with a conserved energy-momentum tensor are obtained via the Lagrange multiplier method.  In Section \ref{sec7} the cosmological implications of the theory are investigated. We discuss and conclude our results in Section \ref{sec8}. We work in a system of units with $c=1$.

\section{The field equations of the $f\left(R,T,R_{\mu\nu}T^{\mu\nu}\right)$ gravity theory}\label{sec2}

We consider that the non-minimal coupling between matter and geometry can be described by the following action,  containing, in addition to the Ricci scalar $R$ and the trace of the energy-momentum tensor $T$, an explicit first  order coupling between the matter energy-momentum $T_{\mu \nu}$ and the
Ricci tensor, respectively,
\be\label{eq200}
S=\f{1}{16\pi G}\int d^4x\sqrt{-g}f\left(R,T,R_{\mu\nu}T^{\mu\nu}\right)+\int
d^4x\sqrt{-g}L_m,
\ee
where $L_m$ is the Lagrangian density of the matter sector, and
the matter energy-momentum tensor $T_{\mu\nu}$ is defined as
\be\label{eq201}
T_{\mu\nu}=-\f{2}{\sqrt{-g}}\f{\delta\left(\sqrt{-g}L_m\right)}{\delta
g^{\mu\nu}}
=g_{\mu\nu}L_m-2\f{\partial L_m}{\partial g^{\mu\nu}}.
\ee
In the second equality we have assumed that the Lagrangian is a function
of the
metric and not its derivatives. The only requirement imposed on the function $f\left(R,T,R_{\mu\nu}T^{\mu\nu}\right)$ is that it  is an arbitrary analytical function in all arguments.

By varying the action given by Eq.~(\ref{eq200}) with respect to the metric we obtain the gravitational field equations as
\begin{widetext}
\begin{align}\label{eq203}
(f_R&-f_{RT}L_m)G_{\mu\nu}+\left[\Box
f_R+\f{1}{2}Rf_R-\f{1}{2}f+f_TL_m+\f{1}{2}\nabla_\alpha\nabla_\beta\left(f_{RT}T^{
\alpha\beta}\right)\right]g_{\mu\nu}-\nabla_\mu\nabla_\nu
f_R+\f{1}{2}\Box\left(f_{RT}T_{\mu\nu}\right)
      \nonumber\\
&+2f_{RT}R_{\alpha(\mu}T_{\nu)}^{~\alpha}-\nabla_\alpha\nabla_{(\mu}\left[T^\alpha_{
~\nu)}f_{RT}\right]
-\left(f_T+\f{1}{2}f_{RT}R+8\pi
G\right)T_{\mu\nu}-2\left(f_Tg^{\alpha\beta}+f_{RT}R^{\alpha\beta}\right)\f{\partial^2
L_m}{\partial g^{\mu\nu}\partial g^{\alpha\beta}}=0.
\end{align}
\end{widetext}

The trace of the gravitational field equation, Eq.~\eqref{eq203}, is obtained as
\begin{eqnarray}\label{trace}
&&3\Box f_R+\f{1}{2}\Box\left(f_{RT}T\right)+\nabla_\alpha\nabla_\beta\left(f_{RT}T^{\alpha\beta}\right)+Rf_R-Tf_T
   \nonumber\\
&&-\f{1}{2}RTf_{RT}+2R_{\alpha\beta}T^{\alpha\beta}f_{RT}+Rf_{RT}L_m +4f_TL_m -2f
   \nonumber\\
&& -8\pi GT -2g^{\mu\nu}\left(g^{\alpha\beta}f_T+R^{\alpha\beta}f_{RT}\right)\f{\partial^2 L_m}{\partial g^{\mu\nu}\partial g^{\alpha\beta}}=0.
\end{eqnarray}

The second derivative of the matter Lagrangian with respect to the metric is non-zero if the matter Lagrangian is the second or of higher order in the metric. Thus, for a perfect fluid with $L_m=-\rho$, or a scalar field with
$L_m=-\partial_\mu\phi\partial^\mu\phi /2$,
this term can be dropped. However, for instance, considering the Maxwell  field, we have
$L_m=-F_{\mu\nu}F^{\mu\nu}/4$, and this term results in
\be\label{eq204-Max}
\f{\partial^2 L_m}{\partial g^{\mu\nu}\partial
g^{\alpha\beta}}=-\f{1}{2}F_{\mu\alpha}F_{\nu\beta},
\ee
thus giving a non-zero contribution to the field equations. In the framework of  $f\left(R,L_m\right)$ theories it has been shown in \cite{soti} that
for a matter source in the form of a perfect fluid,  for a  non-minimally coupled  Ricci scalar and  matter Lagrangian in the form $L_m=p$,
the extra force vanishes in the case of dust. However, in the present case, we will see
that even with this choice,  the extra force does not  vanish  in general.

In analogy with the standard Einstein field equation one can write the gravitational field equation \eqref{eq203} as
\be\label{eq204}
G_{\mu\nu}=8\pi G_{eff}T_{\mu\nu}-\Lambda_{eff}g_{\mu\nu}+T^{eff}_{\mu\nu},
\ee
where we have defined the effective gravitational coupling $G_{eff}$, the effective cosmological constant $\Lambda _{eff}$, and an effective energy-momentum tensor $T^{eff}_{\mu\nu}$ as
\be\label{eq204-1}
G_{eff}=\f{G+\f{1}{8\pi}\big(f_T+\f{1}{2}f_{RT}R-\f{1}{2}\Box f_{RT}\big)}{f_R-f_{RT}L_m},
\ee
\be
\Lambda_{\tiny{eff}}=\f{2\Box
f_R+Rf_R-f+2f_TL_m+\nabla_\alpha\nabla_\beta(f_{RT}T^{
\alpha\beta})}{2(f_R-f_{RT}L_m)},
\ee
and
\bl\label{eq204-2}
T^{eff}_{\mu\nu}&=\f{1}{f_R-f_{RT}L_m}\Bigg\{\nabla_\mu\nabla_\nu f_R-\nabla_\alpha f_{RT}\nabla^\alpha T_{\mu\nu}
    \nonumber\\
&-\f{1}{2}f_{RT}\Box T_{\mu\nu}-
2f_{RT}R_{\alpha(\mu}T_{\nu)}^{~\alpha}+\nabla_\alpha\nabla_{(\mu}\left[T^\alpha_{
~\nu)}f_{RT}\right]
    \nonumber\\
&+2\left(f_Tg^{\alpha\beta}+f_{RT}R^{\alpha\beta}\right)
\f{\partial^2
L_m}{\partial g^{\mu\nu}\partial g^{\alpha\beta}}\Bigg\},
\end{align}
respectively. In general $G_{eff}$ and $\Lambda_{eff}$ are not constants, and they depend on the specific model considered.

It is worth mentioning the main differences between the present theory to that presented in \cite{fRT}. In particular, when assuming a traceless energy-momentum tensor, $T=0$. For instance when the electromagnetic field is involved, the gravitational field equations for the $f(R,T)$ theory reduce to that of the field equations for $f(R)$ gravity and all non-minimal couplings of gravity to the matter field vanish. In contrast, the theory outlined in this work still has a non-minimal coupling to the electromagnetic field via the $R_{\mu\nu}T^{\mu\nu}$ coupling term in the action, which is non-zero in general.

\section{ Equation of motion of the massive test particles in the $f\left(R,T,R_{\mu\nu}T^{\mu\nu}\right)$ gravity theory}\label{sec3}

The covariant divergence of the energy-momentum tensor can be obtained by taking the divergence of the gravitational field equation, Eq.~\eqref{eq203}, which takes the following form
\begin{align}\label{eq401}
&\nabla^\mu T_{\mu\nu}=\f{2}{\left(1+Rf_{TR}+2f_T\right)}\Bigg\{
\nabla_\mu\left(f_{RT}R^{\sigma\mu}T_{\sigma\nu}
\right)
    \nonumber\\
&+\nabla_\nu\left(L_mf_T\right)-\f{1}{2}\bigg(f_{RT}R_{\rho\sigma}+f_T
g_{\rho\sigma}\bigg)
\nabla_\nu T^{\rho\sigma}
     \nonumber\\
 & -G_{\mu\nu}\nabla^\mu \left(f_{RT}L_m\right)-
\f{1}{2}\left[\nabla^\mu\left(Rf_{RT}\right)+2\nabla^\mu f_T\right]T_{\mu\nu}\Bigg\},
\end{align}
where we have assumed that $\partial^2
L_m/\partial g^{\mu\nu}\partial g^{\alpha\beta}=0$, and we have used the mathematical identities
\bea
\nabla^\mu \left(f_R R_{\mu\nu}+\Box f_R
g_{\mu\nu}-\f{1}{2}fg_{\mu\nu}-\nabla_\mu\nabla_\nu
f_R\right)
   \nonumber\\
=-\f{1}{2}\bigg[f_T \nabla_\nu
T+f_{RT}\nabla_\nu\left(R_{\rho\sigma}T^{\rho\sigma}\right)\bigg],
\eea
\be
2T_{\mu\tau;\delta[;\rho;\sigma]}=T_{\mu\tau;\alpha}R^\alpha_{~\delta\rho\sigma}
+T_{\alpha\tau;\delta}R^\alpha_{~\mu\rho\sigma}+T_{\mu\alpha;\delta}R^\alpha_{
~\tau\rho\sigma},
\ee
and $\left[\Box,\nabla_\nu\right]T=R_{\mu\nu}\nabla^\mu T$ respectively.

In order to find the equation of motion for a massive test particle we start with the energy-momentum tensor of the perfect fluid, given by
\begin{align}
T_{\mu\nu}=pg_{\mu\nu}+(\rho+p)u_\mu u_\nu,
  \label{perfectfluid}
\end{align}
where $u^{\mu}$ is the four-velocity of the particle. Taking the divergence of the Eq. (\ref{perfectfluid}), and by introducing the projection operator $h_{\mu \nu }$, defined as
$
h_{\mu \nu }=g_{\mu \nu }+u_{\mu}u_{\nu}
$,
we obtain
\begin{align}
\nabla_\mu T^{\mu\nu}=h^{\mu\nu}\nabla_\mu p &+ u^\nu u_\mu\nabla^\mu \rho\nonumber\\&+(\rho+p)\big(u^\nu\nabla_\mu u^\mu+u^\mu\nabla_\mu u^\nu\big).
\end{align}
Multiplying the above equation with $h_\nu^\lambda$ one finds
\begin{align}
h_\nu^\lambda\nabla_\mu T^{\mu\nu}=(\rho+p)u^\mu\nabla_\mu u^\lambda+h^{\nu\lambda}\nabla_\nu p,\nonumber
\end{align}
where we have used the identity $u_\mu\nabla_\nu u^\mu=0$.
The equation of motion for a massive test particle with the matter Lagrangian $L_m=p$, then takes the form
\be\label{eq404}
\f{d^2x^\lambda}{ds^2}+\Gamma^\lambda_{~\mu\nu}u^\mu u^\nu=f^\lambda,
\ee
where we have used equation \eqref{eq401} to write the covariant divergence of the energy-momentum tensor and the definition of the covariant derivative to obtain the left hand side of the above equation from $u^\mu\nabla_\mu u^\lambda$. The extra force acting on the test particles is given by
\bea\label{eq405}
f^\lambda&=&\f{1}{\rho+p}\Bigg[\left(f_T+Rf_{RT}\right)\nabla_\nu\rho-
\left(1+3f_T\right)\nabla_\nu p\nonumber\\
&-&(\rho+p)f_{RT}R^{\sigma\rho}\left(\nabla_\nu
h_{\sigma\rho}-2\nabla_\rho h_{\sigma\nu}\right)\nonumber\\
&-&f_{RT}
R_{\sigma\rho}h^{\sigma\rho}\nabla_\nu\left(\rho+p\right)\Bigg]
\f{h^{\lambda\nu}}{1+2f_T+Rf_{RT}}.
\eea
Contrary to the nonminimal coupling presented in \cite{extraforcefR}, and as can be seen from the above equations, the extra force does not vanish even with the Lagrangian $L_m=p$.

The extra-force is perpendicular to the four-velocity, satisfying the relation $f^{\lambda }u_{\lambda }=0$. In the absence of any coupling between matter and geometry, with $f_T=f_{RT}=0$, the extra-force takes the usual form of the standard general relativistic fluid motion, i.e., $f^{\lambda }=-h^{\lambda \nu}\nabla _{\nu }p/\left(\rho +p\right)$. In the case of  $f\left(R,T,R_{\mu\nu}T^{\mu\nu}\right)$ gravity theories, there is an explicit dependence of the extra-force on the Ricci tensor $R_{\sigma \rho }$, which makes the deviation from the geodesic motion more important for regions with strong gravitational fields.

\section{The Newtonian limit of $f\left(R,T,R_{\mu\nu}T^{\mu\nu}\right)$ gravity}\label{sec4}

Lets us now consider the Newtonian limit of the theory. Using the weak field and slow motion approximation, we derive the equation of motion of massive test particles in a weak gravitational field as well as the generalized Poisson equation satisfied by the Newtonian potential $\phi $.

\subsection{The equation of motion of massive test particles}

In order to obtain the Newtonian limit, we show first that the equation of motion, Eq.~\eqref{eq404}, can be derived from a variational principle \cite{Harko1, fRT}. To  this end, we assume that one can represent the extra force formally as
\be\label{eq407}
f^{\lambda} = (g^{\nu \lambda}+u^{\nu}u^{\lambda})\nabla_{\nu} \ln \sqrt{Q},
\ee
where $Q$ is a dimensionless function to be determined from the variational principle.
With this assumption, one can prove that the equation of motion
Eq.~\eqref{eq404} can be obtained by varying the action \cite{fRLm}
\begin{align}\label{eq408}
S_p=\int L_p \; ds=\int\sqrt{Q}\sqrt{g_{\mu\nu}u^{\mu}u^{\nu}} \; ds,
\end{align}
where $S_p$ and $L_p$ are the action and  Lagrangian density of the test
particle respectively, provided that $Q$ is not an
explicit function of $u^{\mu}$.
When $\sqrt{Q}\rightarrow1$, we obtain the variational principle for the standard general relativistic  motion for a massive test particle.

In order to obtain the function $Q$ for Eq.~\eqref{eq404} in the Newtonian
limit, we assume that the density of the physical system is small and therefore the pressure satisfies the condition $p \ll \rho $. Hence the energy-momentum tensor of the system can be taken as the energy-momentum tensor
of pressureless dust. Moreover, by considering the limiting case of small velocities, we can take the four-velocity in the form
$u^\mu=\delta^\mu_0/\sqrt{g_{00}}$ and  drop the covariant derivatives
of $h_{\mu\nu}$ in
Eq.~\eqref{eq405}. Therefore  Eq.~\eqref{eq404} takes the form
\be\label{eq412}
f^\lambda=\f{F}{\rho}h^{\lambda\nu}\nabla_\nu\rho,
\ee
where
\be\label{eq413}
F=\f{f_T+f_{RT}(R-R_{\alpha\beta}h^{\alpha\beta})}{1+2f_T+Rf_{RT}}.
\ee
We also note that $F$ is dimensionless.
In the Newtonian limit, one can expand the energy density around the background energy density $\rho_0$ as
$
\rho=\rho_0+\delta\rho
$
and then the function $F$ can be expanded as
\be\label{eq}
F(\rho)=F(\rho_0)+\left.\f{dF}{d\rho}\right|_{\rho_0}\left(\rho-\rho_0\right)\equiv
F_0+F_1\delta\rho.
\ee
where we have denoted $\delta\rho\equiv\rho-\rho_0$. The expression \eqref{eq412} can then be expanded in the first order in $\delta\rho$ as
\begin{align}\label{}
\f{F}{\rho}\nabla_\nu\rho \approx F_0\nabla_\nu\delta,
\end{align}
where we define fractional energy density perturbation as $\delta=\delta\rho/\rho_0$.
From the expression above one can read off the dimensionless quantity $\sqrt{Q}$ for small $\rho$ as
\be\label{eq414}
\sqrt{Q}\approx1+F_0\delta = 1-F_0+\f{F_0}{\rho_0}\rho,
\ee
We have therefore obtained $\sqrt{Q}$  in the case of dust as an explicit
function of the energy density $\rho$.
We may now proceed to study the Newtonian limit of the theory by using the variational principle Eq.~\eqref{eq408}, and also Eq.~\eqref{eq414}. In the weak field limit the interval $ds$ for dust moving in a gravitational field is
\be\label{eq415}
ds\approx \sqrt{1+2\phi-\vec{v}^2} \; dt\approx \left(1+\phi-\frac{\vec{v}^2}{2}\right) \, dt,
\ee
where $\phi$ is the Newtonian potential and $\vec{v}$ is the three-dimensional
velocity of the fluid.
The equation of motion of the fluid to first order approximation can
be obtained from the variational principle
\be\label{eq417}
\delta\int \left[1+U(\rho)+\phi-\frac{\vec{v}^2}{2}\right]dt=0.
\ee
The total acceleration of the system, $\vec{a}$, is given as
\be\label{eq418}
\vec{a}=-\vec{\nabla} \phi-\vec{\nabla} U(\rho)=\vec{a}_N+\vec{a}_E,
\ee
where $\vec{a}_N=-\vec{\nabla} \phi$ is the Newtonian acceleration, and the
supplementary acceleration, induced by the geometry-matter coupling, is
\be\label{eq419}
\vec{a}_E(\rho)=-\vec{\nabla} U(\rho)=\f{F_0}{\rho_0}\vec{\nabla}
\rho.
\ee

The acceleration given by Eq.~(\ref{eq419}) is due to the modification of the gravitational action.
In our case, there is no hydrodynamical acceleration $\vec{a}_p$ term in the
total acceleration, because of our assumption that the fluid is pressureless. However, such an acceleration
does exist in the general case. We see from Eq.~(\ref{eq419}) that the extra
acceleration $\vec{a}_E$ is essentially due to the non-minimal coupling between matter and geometry. The extra-acceleration is
proportional to the gradient of the energy density of the fluid. Therefore,  for a
constant energy density source  and a pressureless fluid,  the extra acceleration vanishes.

\subsection{The generalized Poisson equation}\label{poi}

To obtain the  Poisson equation we assume that the matter content of the self-gravitating system is represented by dust. Also,  noting that in the Newtonian limit one has $R=-2R_{00}=-2\nabla^2\phi$, where $\phi$ is the Newtonian potential which appears in the $(00)$ component of the metric $g_{00}=-(1+2\phi)$, one can compute the individual terms in the trace equation \eqref{trace} as
\be
R_{\alpha\beta}T^{\alpha\beta}\sim \rho\nabla^2\phi,\nonumber
\ee
\be
\Box f_R\sim\nabla^2 f_R+\nabla f_R \cdot \nabla\phi,\nonumber
\ee
and
\be
\nabla_\alpha\nabla_\beta(f_{RT}T^{\alpha\beta})\sim \nabla(\rho f_{RT}) \cdot \nabla\phi+\rho f_{RT}\nabla^2\phi,\nonumber
\ee
respectively.

Substituting the above expressions into Eq.~\eqref{trace} and rearranging terms, we obtain the generalized Poisson equation as
\bea\label{poison}
\nabla^2\phi&=&\f{1}{2(f_R-2\rho f_{RT})}\bigg[8\pi G\rho+3\nabla^2 f_R-3\rho f_T\nonumber\\
& -&2f+ \nabla(3f_R+\rho f_{RT}) \cdot \nabla \phi\bigg].
\eea
As can be seen, the generalized Poisson equation is modified by the addition of
gradient of the $\phi$ field to the equation.

\section{The Dolgov-Kawasaki instability in $f\left(R,T,R_{\mu\nu}T^{\mu\nu}\right)$ gravity}\label{sec5}
Beside consistency with the Solar System tests, any gravitational theory should be stable against classical and quantum fluctuations. One of the important instabilities of modified theories of gravity is the Dolgov-Kawasaki instability \cite{DK, DK1}, which we shall discuss in the present Section.

Let us assume that, in order to be consistent with the Solar System tests, the Lagrangian can be written as
\be
f(R,T,R_{\mu\nu}T^{\mu\nu})=R+\epsilon\Phi\left(R,T,R_{\mu\nu}T^{\mu\nu}\right),
 \ee
where $\epsilon$ is a small parameter. Following \cite{DK}, we expand the space-time quantities around a constant curvature background with geometrical and physical parameters $\left(\eta _{\mu \nu},R_0, T_{\mu \nu }^0, T_0,L_0\right)$, so that
\begin{align}
R_{\mu\nu}&=\f{1}{4}R_0\eta_{\mu\nu}+R^1_{\mu\nu},  \qquad  R=R_0+R_1,\nonumber\\
T_{\mu\nu}&=T^0_{\mu\nu}+T^1_{\mu\nu},  \qquad  T=T_0+T_1,\nonumber\\
 L_m&=L_0+L_1,
\end{align}
where we have locally expanded the metric tensor as
$
g_{\mu\nu}=\eta_{\mu\nu}+h_{\mu\nu}
$.
We note that in the above equations we have really two types of approximations, as mentioned in \cite{DK1}. The first is an adiabatic expansion around a constant curvature space, which is justified on the time-scales much shorter than the Hubble time. The second approximation is a local expansion in the small regions of space-time, which are locally flat. These approximations have been used extensively in $f(R)$ gravity theories, \cite{DK,DK1}.
The function $f\left(R,T,R_{\mu\nu}T^{\mu\nu}\right)$ can be expanded as
\begin{align}
f(R,&T,R_{\mu\nu}T^{\mu\nu})
=R_0+R_1+\epsilon\bigg[\Phi(0)+\Phi_R(0)R_1
   \nonumber\\
&+\Phi_T(0)T_1+\Phi_{RT}(0)\left(\f{1}{4}R_0T^1+R^1_{\mu\nu}T_0^{\mu\nu}\right)\bigg]\nonumber\\
&=R_0+\epsilon\Phi(0)+\big[1+\epsilon\Phi_R(0)\big]R_1+H^{(1)},
\end{align}
where $(0)$ denotes the computation of the function at the background level, and for simplicity we have defined the first order quantity $H^{(1)}$ as
 \be
 H^{(1)}=\epsilon\left[\Phi_T(0)T_1+\Phi_{RT}(0)\left(\f{1}{4}R_0T^1+R^1_{\mu\nu}T_0^{\mu\nu}\right)\right].
 \ee
 We then obtain
\bea
f_R=1+\epsilon\Phi_R(0)+\epsilon\Phi_{R,R}(0)R_1+H_R^{(1)},  \\
f_T=\epsilon\Phi_T(0)+\epsilon\Phi_{T,R}(0)R_1+H_T^{(1)},  \\
f_{RT}=\epsilon\Phi_{RT}(0)+\epsilon\Phi_{RT,R}(0)R_1+H_{RT}^{(1)}.
\eea
The trace equation \eqref{trace} can then be expanded to first order to obtain
\begin{widetext}
\bea
&&\bigg(3\epsilon\Phi_{R,R}(0)+\f{1}{2}\epsilon T_0\Phi_{RT,R}(0)\bigg)\Box R_1+\epsilon T_0^{\alpha\beta}\Phi_{RT,R}(0)\nabla_\alpha\nabla_\beta R_1+
\bigg[f_R(0)+\epsilon R_0\Phi_{R,R}(0)-\epsilon T_0\Phi_{T,R}(0)  -\f{1}{2}T_0f_{RT}(0)
   \nonumber\\
&&-\f{1}{2}\epsilon R_0T_0\Phi_{RT,R}(0)
+\f{1}{2}\epsilon R_0T_0\Phi_{RT,R}(0)+\epsilon R_0L_0\Phi_{RT,R}(0)+f_{RT}(0)L_0+4\epsilon L_0\Phi_{T,R}(0)-2-\epsilon\Phi_R(0)\bigg]R_1
    \nonumber\\
&&+3\Box H_R^{(1)}+\f{1}{2}f_{RT}(0)\Box T_1+\f{1}{2}T_0\Box H_{RT}^{(1)}+f_{RT}(0)\nabla_\alpha\nabla_\beta T_1^{\alpha\beta}+T_0^{\alpha\beta}\nabla_\alpha\nabla_\beta H_{RT}^{(1)}+R_0 H_R^{(1)}-T_1 f_T(0)      \nonumber\\
&&-T_0 H_T^{(1)}+2R^1_{\mu\nu}T_0^{\mu\nu}f_{RT}(0)+R_0f_{RT}(0)L_1+R_0L_0 H_{RT}^{(1)}+4f_T(0)L_1+4L_0 H_T^{(1)}
   \nonumber\\
&& -2 H^{(1)}+8\pi GT_1-2\eta^{\mu\nu}\eta^{\alpha\beta}\big[f_T(0)+\f{1}{4}R_0f_{RT}(0)\big]\f{\partial^2 L_1}{\partial g^{\mu\nu}\partial g^{\alpha\beta}}=0.
\eea

In the limit considered, one may write $\Box=-\partial_t^2+\nabla^2$, thus obtaining
\be\label{}
T_0^{\alpha\beta}\nabla_\alpha\nabla_\beta R_1=T^{00}_0 \ddot{R}_1+T_0^{ij}\partial_i\partial_j R_1.
\ee
One can then rewrite the above equation as
\be
\ddot{R}_1+V_{eff}^{ij}\nabla_i\nabla_j R_1+m_{eff}^2 R_1 = H_{eff},
\ee
where we have defined
\be
V_{eff}^{ij}=\f{\big(3\epsilon\Phi_{R,R}(0)+\f{1}{2}\epsilon T_0\Phi_{RT,R}(0)\big)\delta^{ij}+\epsilon T_0^{ij}\Phi_{RT,R}(0)}{T_0^{00}-3\epsilon\Phi_{R,R}(0)-\f{1}{2}\epsilon T_0\Phi_{RT,R}(0)},
\ee
and
\bea
H_{eff}&=&\big[3\epsilon\Phi_{R,R}(0)+\f{1}{2}\epsilon T_0\Phi_{RT,R}(0)-T_0^{00}\big]^{-1}\bigg\{ 3\Box H_R^{(1)}+\f{1}{2}f_{RT}(0)\Box T_1+\f{1}{2}T_0\Box H_{RT}^{(1)}+f_{RT}(0)\nabla_\alpha\nabla_\beta T_1^{\alpha\beta}\nonumber\\
&&+T_0^{\alpha\beta}\nabla_\alpha\nabla_\beta H_{RT}^{(1)}+R_0 H_R^{(1)}-T_1 f_T(0)
-T_0 H_T^{(1)}+2R^1_{\mu\nu}T_0^{\mu\nu}f_{RT}(0)+R_0f_{RT}(0)L_1  + R_0L_0 H_{RT}^{(1)}
  \nonumber\\
&&+4f_T(0)L_1+4L_0 H_T^{(1)}-2 H^{(1)}+8\pi GT_1-2\eta^{\mu\nu}\eta^{\alpha\beta}\bigg[f_T(0)+\f{1}{4}R_0f_{RT}(0)\bigg]\f{\partial^2 L_1}{\partial g^{\mu\nu}\partial g^{\alpha\beta}}\bigg\},
\eea
respectively, and we have introduced  the effective mass  $m_{eff}$ as
\bea
m_{eff}^2&=&\left[\big(T_0^{00}-\f{1}{2} T_0\big) f_{RT,R}(0)-3f_{RR}(0)\right]^{-1}
\bigg[ \epsilon R_0\Phi_{R,R}(0)-\epsilon T_0\Phi_{T,R}(0)-\f{1}{2}\epsilon R_0T_0\Phi_{RT,R}(0)
\nonumber\\
&&- \f{1}{2}\epsilon T_0\Phi_{RT}(0)+
\f{1}{2}\epsilon R_0T_0\Phi_{RT,R}(0)+
\epsilon R_0L_0\Phi_{RT,R}(0)+\epsilon\Phi_{RT}(0)L_0+
4\epsilon L_0\Phi_{T,R}(0)-1-\epsilon\Phi_R(0)\bigg].
\eea
\end{widetext}

The dominant term in the above expression is $1/\big[3f_{RR}(0)+(\f{1}{2}T_0 -T_0^{00})f_{RT,R}(0)\big]$, and therefore the condition to avoid the Dolgov-Kawasaki instability is
\be
3f_{RR}(0)-\left(\rho_0-\f{1}{2}T_0 \right)f_{RT,R}(0)\geq0,
\ee
where $\rho _0$ is the background energy density of the matter \cite{DK}. We note that due to the above expression, condition for the stability does not depend on the derivative of the function $f$ with respect to $T$. So, the DK stability condition for the case of $f(R,T)$ gravity is the same as  $f(R)$ gravity. However, the condition is modified in the case of $f(R,T,R_{\mu\nu}T^{\mu\nu})$.


\section{$f\left(R,T,R_{\mu\nu}T^{\mu\nu}\right)$ gravity theories with energy-momentum conservation}\label{sec6}

The general non-minimal coupling between matter and geometry leads to the important consequence that the matter energy-momentum tensor is not conserved. In Section \ref{sec3} we have shown that this property of the gravitational  theory  determines the appearance of the extra force. However, the energy non-conservation can be interpreted as a shortcoming of these types of theories. In the framework of the $f(R,T)$ theory, models with energy conservation have been investigated in \cite{Gom}. By assuming a specific additive form for the function $f(R,T)$, $f(R,T)=f_1(R)+f_2(T)$, and by imposing the condition of the energy conservation, and under the assumption of a barotropic fluid, the function $f_2(T)$ can be uniquely determined as $f_2(T)\sim T^{1/2}$. In the following we investigate the energy conservation in $f\left(R,T,R_{\mu\nu}T^{\mu\nu}\right)$ gravity.

In order to impose the matter energy-momentum tensor conservation, one can use the Lagrange multiplier method \cite{odi1}. To do effect, let us consider the modified action
\bea
S&=&\f{1}{16\pi G}\int d^4x\sqrt{-g}\bigg[f\left(R,T,R_{\mu\nu}T^{\mu\nu}\right)+\lambda^\mu\nabla^\nu T_{\mu\nu}\bigg]\nonumber\\
&&+ \int d^4x\sqrt{-g}L_m,
\eea
where we have introduced the vector Lagrange multiplier $\lambda^\mu$. The variations of the first and the third terms are similar to those computed in Section~\ref{sec2}. The variation of the second term with respect to the metric is given by
\begin{align}
&\delta(\sqrt{-g}\lambda^\mu\nabla^\nu T_{\mu\nu})\nonumber\\&=\sqrt{-g}\bigg[\lambda^\alpha(\nabla_\mu T_{\alpha\nu}-\f{1}{2}\nabla^\beta T_{\alpha\beta}g_{\mu\nu})\delta g^{\mu\nu}+\lambda^\mu\nabla^\nu\delta T_{\mu\nu}\bigg],
\end{align}
where the variation of the energy-momentum tensor is obtained from Eq.~\eqref{eq201}. Combining  the above results with the calculations of Section \ref{sec2}, we obtain the field equations together with the
energy-momentum conservation as
\begin{widetext}
\bea\label{lf}
&&\left(f_R-f_{RT}L_m\right)G_{\mu\nu}+\left[\Box
f_R+\f{1}{2}Rf_R-\f{1}{2}f+f_TL_m+\f{1}{2}\nabla_\alpha\nabla_\beta\left(f_{RT}T^{
\alpha\beta}\right)\right]g_{\mu\nu}-\nabla_\mu\nabla_\nu
f_R+\f{1}{2}\Box(f_{RT}T_{\mu\nu})
     \nonumber\\
&&+2f_{RT}R_{\alpha(\mu}T_{\nu)}^{~\alpha}-\nabla_\alpha\nabla_{(\mu}\left[T^\alpha_{
~\nu)}f_{RT}\right]
-(f_T+\f{1}{2}f_{RT}R+8\pi
G)T_{\mu\nu}-2(f_Tg^{\alpha\beta}+f_{RT}R^{\alpha\beta})\f{\partial^2
L_m}{\partial g^{\mu\nu}\partial g^{\alpha\beta}}
     \nonumber\\
&&-\f{1}{2}\lambda^\alpha\nabla^\rho T_{\alpha\rho}g_{\mu\nu}+\lambda_\rho\nabla_{(\mu}T^\rho_{\nu)}-\nabla_{(\mu}\lambda_{\nu)}L_m-\f{1}{2}\nabla_\alpha\lambda^\alpha\big(L_m g_{\mu\nu}-T_{\mu\nu}\big)+2\nabla^{(\alpha}\lambda^{\beta)}\f{\partial^2
L_m}{\partial g^{\mu\nu}\partial g^{\alpha\beta}}
=0.
\eea

Now, variation with respect to the vector $\lambda^\mu$ results in
\be\label{encons}
\nabla^\nu T_{\mu\nu}=0,
\ee
which is the conservation of the energy-momentum tensor. Therefore Eqs.~(\ref{lf}) and (\ref{encons}) provide the basic equations of the $f\left(R,T,R_{\mu\nu}T^{\mu\nu}\right)$ gravity theory with energy conservation. The gravitational field equations explicitly depend on the Lagrange multiplier $\lambda ^{\mu }$. The field equations  in the case of the matter Lagrangian $L_m=-\rho$ or $L_m=p$ which leads to $\partial^2 L_m/\partial g^{\mu\nu}\partial g^{\alpha\beta}\equiv 0$ take a simpler form.
For $L_m=-\rho$ we obtain
 \begin{eqnarray}
(f_R+\rho f_{RT})G_{\mu\nu}+\big[\Box f_R+\f{1}{2}Rf_R-\f{1}{2}f-\rho f_T+\f{1}{2}T^{\alpha\beta}\nabla_\alpha\nabla_\beta f_{RT}\big]g_{\mu\nu}-\nabla_\mu\nabla_\nu f_R+\f{1}{2}\Box(f_{RT}T_{\mu\nu})+2f_{RT}R_{\alpha(\mu}T_{\nu)}^{~\alpha}
      \nonumber\\
-\nabla_\alpha\nabla_{(\mu}\big[T^\alpha_{~\nu)}f_{RT}\big]-(f_T+\f{1}{2}f_{RT}R+8\pi G)T_{\mu\nu}+\lambda_\rho\nabla_{(\mu}T^\rho_{~\nu)}+\rho\nabla_{(\mu}\lambda_{\nu)}+\f{1}{2}(\rho g_{\mu\nu}+T_{\mu\nu})\nabla_\alpha\lambda^\alpha=0,
\end{eqnarray}
\end{widetext}
 where the conservation of the energy-momentum tensor is taken into account.

In the case of the electromagnetic field, because the trace of the energy-momentum vanishes, we have $f_T=0$, and using Eq.~\eqref{eq204-Max} we find the field equations
 \begin{widetext}
 \begin{align}
(f_R+\f{1}{4}F^2f_{RT})G_{\mu\nu}&+\big[\Box f_R+\f{1}{2}Rf_R-\f{1}{2}f-\rho f_T+\f{1}{2}T^{\alpha\beta}\nabla_\alpha\nabla_\beta f_{RT}\big]g_{\mu\nu}-\nabla_\mu\nabla_\nu f_R+\f{1}{2}\Box(f_{RT}T_{\mu\nu})+2f_{RT}R_{\alpha(\mu}T_{\nu)}^{~\alpha}\nonumber\\&-\nabla_\alpha\nabla_{(\mu}\big[T^\alpha_{~\nu)}f_{RT}\big]-(\f{1}{2}f_{RT}R+8\pi G)T_{\mu\nu}+f_{RT}R^{\alpha\beta}F_{\mu\alpha}F_{\nu\beta}+\lambda_\rho\nabla_{(\mu}T^\rho_{~\nu)}+\f{1}{4}F^2\nabla_{(\mu}\lambda_{\nu)}\nonumber\\&+\f{1}{2}F_{\mu\nu}^2\nabla_\alpha\lambda^\alpha-\nabla^{(\alpha}\lambda^{\beta)}F_{\mu\alpha}F_{\nu\beta}=0,
\end{align}
\end{widetext}
where we have defined $F^2=F_{\alpha\beta}F^{\alpha\beta}$ and $F^2_{\mu\nu}=F_{\mu\alpha}F_\nu^{~\alpha}$ and used the conservation of the energy-momentum tensor.

\section{Cosmological applications of $f\left(R,T,R_{\mu\nu}T^{\mu\nu}\right)$ gravity}\label{sec7}

Let us now consider some examples of cosmological solutions of the theory. In sections \ref{A}-\ref{C} we will consider the cosmology of the standard theory without the energy-momentum conservation, and in section \ref{D} we will consider the cosmology of the conservative case. In order to obtain explicit results and as a first step, one has to fix the functional form of the function $f\left(R,T,R_{\mu\nu}T^{\mu\nu}\right)$. In the following we  consider three specific choices for $f$, namely $f=R+\alpha R_{\mu\nu}T^{\mu\nu}$, $f=R+\alpha R_{\mu\nu}T^{\mu\nu}+\beta\sqrt{T}$ and $f=R+\alpha R R_{\mu\nu}T^{\mu\nu}$, where $\alpha,~\beta ={\rm constant}$, respectively. We analyze the evolution and  dynamics of the Universe for the above with and without energy conservation. In all cases we assume that the Universe is isotropic and homogeneous, with the matter content described by the energy density $\rho $, and thermodynamic pressure $p$ with the matter Lagrangian as $L_m=-\rho$. The geometry of the space-time
is described by the Friedmann-Lemaitre-Robertson-Walker (FLRW) metric, given by
\be
ds^2=-dt^2+a^2(t)(dx^2+dy^2+dz^2),
\ee
where $a(t)$ is the scale factor of the Universe. We define the Hubble parameter as $H=\dot{a}/a$, and we describe the accelerated expansion of the Universe through the values of the deceleration parameter $q$, defined as
\be
q=\frac{d}{dt}\frac{1}{H}-1.
\ee
If $q<0$, the expansion of the Universe is accelerating, while positive values of $q$, $q\geq 0$, describe decelerating evolutions.

\subsection{Specific case I: $f=R+\alpha R_{\mu\nu}T^{\mu\nu}$}\label{A}

Let us first consider the simplest case, in which the interaction between matter and geometry takes place only via the coupling between the energy-momentum and  Ricci tensors. This simple case can also show the main differences of the present  theory with the so-called $f(R,T)$ gravity theory \cite{fRT}.
The gravitational field equations for this form of $f$ are given by
\begin{align}\label{eq203-1}
G_{\mu\nu}&+\alpha\Bigg[2R_{\sigma(\mu}T^\sigma_{~\nu)}-
\f{1}{2}R_{
\rho\sigma}T^{\rho\sigma}g_{\mu\nu}-\f{1}{2}RT_{\mu\nu}\nonumber\\
&-\f{1}{2}\left(2 \nabla_\sigma\nabla_{(\nu}
T^\sigma_{~\mu)}-\Box
T_{\mu\nu}-\nabla_\alpha\nabla_\beta T^{\alpha\beta}g_{\mu\nu}\right)\nonumber\\
&-G_{\mu\nu}L_m-2R^{\alpha\beta}\f{\partial^2 L_m}{\partial g^{\mu\nu}\partial
g^{\alpha\beta}}\Bigg]-
8\pi GT_{\mu\nu}=0.
\end{align}

The effective gravitational coupling, the effective cosmological constant, and the effective energy-momentum tensor are given for this choice of $f$ by
\bea
G_{eff}&=&\f{16\pi G+\alpha R}{16\pi(1-\alpha L_m)}, \label{eq205} \\
\Lambda_{eff}&=&\f{\alpha}{2(1-\alpha L_m)}(\nabla_\alpha\nabla_\beta-R_{\alpha\beta})T^{\alpha\beta},
\eea
and
\bea\label{eq206}
T^{eff}_{\mu\nu}&=&\f{\alpha}{2(1-\alpha L_m)}\bigg[g_{\mu\beta}\nabla_\alpha\nabla_\nu+g_{\beta\nu}\nabla_\alpha\nabla_\mu \nonumber\\
&&-g_{\mu\alpha}g_{\nu\beta}\Box-
4R_{\alpha(\mu}g_{\nu)\beta}\bigg]T^{\alpha\beta}.
\eea

For the case of the FLRW metric the independent cosmological field equations are
\be\label{c1}
3H^2=\frac{\kappa}{1-\alpha \rho }\rho +\frac{3}{2}\frac{\alpha }{1-\alpha \rho }H\left(\dot{\rho }-\dot{p}\right),
\ee
and
\be\label{c2}
2\dot{H}+3H^2=\frac{2\alpha }{1+\alpha p}H\dot{\rho }-\frac{\kappa p}{1+\alpha p}+\frac{1}{2}\frac{\alpha }{1+\alpha p}\left(\ddot{\rho }-\ddot{p}\right),
\ee
respectively, where we have denoted $\kappa =8\pi G$ for simplicity. When $\alpha =0$ we recover the standard Friedmann equations. To remove the under determinacy of the field equations, we must impose an equation of state for the cosmological matter, $p=p(\rho)$. A standard form of the cosmological matter equation of state is $p=\omega \rho $, where $\omega ={\rm constant}$, and $0\leq \omega \leq 1$.

\subsubsection{High cosmological density limit of the field equations}

We shall first consider the high energy density limit of the system of modified cosmological equations (\ref{c1}) and (\ref{c2}). Moreover, we assume that the constant $\alpha $ is small, so that $\alpha \rho \ll 1$, and $\alpha p \ll  1$, respectively. In the high-energy limit, $\rho =p$, and Eqs.~(\ref{c1}) and (\ref{c2}) take the approximate form
\bea
3H^2&=&\kappa \rho,
   \\
2\dot{H}+3H^2&=&-\kappa \rho +2\alpha H\dot{\rho }.
\eea

The time evolution of the Hubble parameter is described by the equation
\be
\left(1-\frac{6\alpha}{\kappa }H^2\right)\dot{H}+3H^2=0,
\ee
and hence for this model the evolution of the Hubble parameter is given by
\be
H(t)= \frac{\sqrt{\left(C_1+3 \kappa  t\right){}^2-24 \alpha  \kappa }+C_1+3 \kappa  t}{12
   \alpha },
\ee
where $C_1$ is an integration constant. One can see that $\alpha>0$ in order to have a positive Hubble parameter. The scale factor of the Universe is given by
\be
a(t)=C_2\frac{\exp \left[\frac{\left(C_1+3 \kappa  t\right) \sqrt{\left(C_1+3 \kappa
   t\right){}^2-24 \alpha  \kappa }+9 \kappa  t^2+6 \kappa C_1 t}{72 \alpha
   \kappa }\right]}{\sqrt[3]{\sqrt{\left(C_1+3 \kappa  t\right){}^2-24 \alpha  \kappa
   }+C_1+3 \kappa  t}},
   \ee
where $C_2$ is an integration constant. In order to have a positive scale factor one should impose that $C_2>0$. In order to have a physical solution, the scale factor should be real for all times including the $t=0$. So one may impose the following constraint on $C_1$
\be
C_1\geq \sqrt{24 \kappa \alpha}.
\ee

The values of the integration constant can be determined from the condition $H(0)=H_0$, and $a(0)=a_0$, where $H_0$ and $a_0$ are the initial values of the Hubble parameter and of the scale factor of the Universe, respectively. This condition immediately provides for $C_1$ the following value
\be\label{c1b}
C_1=\frac{6 \alpha  H_0^2+\kappa }{H_0}.
   \ee
  For the integration constant $C_2$ we obtain
  \bea
  C_2&=&a_0\sqrt[3]{\sqrt{\frac{\left(\kappa -6 \alpha  H_0^2\right)^2}{H_0^2}}+6 \alpha
   H_0+\frac{\kappa }{H_0}} \times \nonumber\\
   &&\times \exp \left[-\frac{\sqrt{\frac{\left(\kappa -6 \alpha
   H_0^2\right)^2}{H_0^2}} \left(6 \alpha  H_0^2+\kappa \right)}{72 \alpha  H_0 \kappa
   }\right].
  \eea

In the small time limit, the scale factor can be represented by
\be
a(t)\approx a_0\left(1+\frac{\kappa }{6H_0\alpha}t\right).
\ee

The deceleration parameter is obtained as
\begin{widetext}
\be
q(t)=-\frac{36 \alpha  H_0 \kappa }{\sqrt{\frac{\left(6 \alpha  H_0^2+3 H_0 \kappa  t+\kappa
   \right)^2}{H_0^2}-24 \alpha  \kappa } \left[6 \alpha  H_0^2+H_0 \sqrt{\frac{\left(6
   \alpha  H_0^2+3 H_0 \kappa  t+\kappa \right)^2}{H_0^2}-24 \alpha  \kappa }+3 H_0 \kappa
   t+\kappa \right]}-1,
   \ee
   \end{widetext}
   and it can be represented in a form of a power series as
   \be
   q(t)\approx -1-\frac{18 \alpha  H_0^2
   }{\kappa -6
   \alpha  H_0^2}+
   \frac{6 H_0 \kappa ^2 }{\left(\kappa -6 \alpha  H_0^2\right)^3}t.
    \ee
For small values of time, if  $24\alpha H_0^2 \ll \kappa$, $q\approx -1$, and the Universe starts its expansion from a de Sitter like phase, entering, after a finite time interval, into a decelerating phase. On the other hand, if $\kappa >6 \alpha  H_0^2$, $q<-1$, and the non-singular Universe experiences an initial super-accelerating phase.

\subsubsection{The case of dust matter}

Next we consider the case of low density cosmological  matter, with $p=0$. Moreover, we assume again that the condition $\alpha \rho  \ll 1$ holds. Then the gravitational field equations, Eqs.~(\ref{c1}) and (\ref{c2}), corresponding to a FLRW Universe, take the approximate form
    \be\label{c3}
    3H^2=\kappa \rho +\frac{3}{2}\alpha H\dot{\rho },
    \ee
    \be\label{c4}
    2\dot{H}+3H^2=2\alpha H\dot{\rho }+\frac{1}{2}\alpha \ddot{\rho }.
    \ee

First we consider the matter dominated phase of the model, in which the non-accelerating expansion of the Universe can be described by a power law form of the scale factor, so that $a=t^m$, $m={\rm constant}$, and $H=m/t$, respectively. The deceleration parameter is given by $q=1/m-1$ Therefore Eq.~(\ref{c3}) gives for the time evolution of the density the equation
  \be
  \frac{3\alpha m}{2t}\dot{\rho }+\kappa \rho -3\frac{m^2}{t^2}=0,
  \ee
  with the general solution given by
  \be
  \rho (t)=\frac{e^{-\frac{\kappa  t^2}{3 \alpha }} \left[3 \rho _0 \alpha  e^{\frac{\kappa
   t_0^2}{3 \alpha }}+\text{Ei}\left(\frac{t^2 \kappa }{3 \alpha
   }\right)-\text{Ei}\left(\frac{t_0^2 \kappa }{3 \alpha }\right)\right]}{3
   \alpha },
   \ee
where $\text{Ei}(z)=-\int_{-z}^{\infty}{e^{-t}dt/t}$ is the exponential integral function, and we have used the initial condition $\rho \left(t_0\right)=\rho _0$. By substituting the expressions of the density and of the Hubble parameter into Eq.~(\ref{c4}), to first order, we obtain the following constraint on $m$,
   \be
   \frac{9 m^2-10 m+1}{3t^2}+O\left(t^2\right)\approx 0,
   \ee
which is (approximately) satisfied if $m$ is given by the algebraic equation $9m^2-10m+1=0$, having the solutions $m_1=1$, and $m_2=1/9$, respectively. The deceleration parameters corresponding to these solutions are $q_1=0$, and $q_2=8$, respectively.  Since a value of the deceleration parameter of the order of $q=8$ seems to be ruled out by the observations, the physical solution has a scale factor $a=t$, and $q=0$. The cosmological solutions with zero value of the deceleration parameter are called marginally accelerating, and they describe the pre-accelerating phase of the cosmic expansion.

Now we look for a de Sitter-type solution of the field equations for the pressureless matter, Eqs.~(\ref{c3}) and (\ref{c4}), by taking $H=H_0={\rm constant}$. Then it follows that, in order to have an accelerated expansion, the matter density must satisfy the equation
   \be
   \ddot{\rho }-H_0\dot{\rho }+\frac{2\kappa }{\alpha }\rho =0,
   \ee
with the general solution given by
   \bea\label{r1}
   \rho (t)&=&e^{\frac{1}{2} H_0 \left(t-t_0\right)} \times \nonumber\\
   &&\Bigg\{\frac{\sqrt{\alpha } \left(2 \rho
   _{01}-H_0 \rho _0\right) }{\sqrt{\alpha  H_0^2-8
   \kappa }}\sinh \left[\frac{ \sqrt{\alpha
  H_0^2-8 \kappa }}{2 \sqrt{\alpha }}\left(t-t_0\right)\right]\nonumber\\
   &&+\cosh \left[\frac{ \sqrt{\alpha  H_0^2-8 \kappa }}{2
   \sqrt{\alpha }}\left(t-t_0\right)\right]\Bigg\},
   \eea
where we have used the initial conditions $\rho \left(t_0\right)=\rho _0$, and $\dot{\rho }\left(t_0\right)=\rho _{01}$, respectively.
  Therefore, in the presence of a non-trivial geometry-matter coupling, once the evolution of the matter density is given by Eq.~(\ref{r1}), the time evolution of the Universe is of the de Sitter type.

\subsection{Specific case II: $f=R+\beta\sqrt{|T|}+\alpha  R_{\mu \nu}T^{\mu \nu }$}\label{B}

In this section, we generalize the previous action by adding a term $\beta\sqrt{T}$, $\beta ={\rm constant}$. Such a model, satisfying the energy conservation, was considered, in the framework of the $f(R,T)$ theory, in \cite{Gom}, where a model with action given by $f(R,T)=R+\beta T^{1/2}$ was investigated.  The field equations of the $f\left(R,T,R_{\mu\nu}T^{\mu\nu}\right)$ gravity in this case are
\begin{align}\label{field1}
G_{\mu\nu}&+\alpha\Bigg[2R_{\sigma(\mu}T^\sigma_{~\nu)}-
\f{1}{2}R_{
\rho\sigma}T^{\rho\sigma}g_{\mu\nu}-\f{1}{2}RT_{\mu\nu}\nonumber\\
&-\f{1}{2}\left(2 \nabla_\sigma\nabla_{(\nu}
T^\sigma_{~\mu)}-\Box
T_{\mu\nu}-\nabla_\alpha\nabla_\beta T^{\alpha\beta}g_{\mu\nu}\right)\nonumber\\
&-G_{\mu\nu}L_m-2R^{\alpha\beta}\f{\partial^2 L_m}{\partial g^{\mu\nu}\partial
g^{\alpha\beta}}\Bigg]-
8\pi GT_{\mu\nu}\nonumber\\&+\f{\beta}{2\sqrt{T}}\bigg[(L_m-T)g_{\mu\nu}-T_{\mu\nu}-2g^{\alpha\beta}\f{\partial^2 L_m}{\partial g^{\mu\nu}\partial
g^{\alpha\beta}}\bigg]=0.
\end{align}
The cosmological equations of this model with the perfect fluid matter in the FRW space-time can be written as
\begin{align}\label{f2}
3(1-\alpha \rho )H^2=\kappa\rho +\frac{3}{2}\alpha H\left(\dot{\rho }-\dot{p}\right)-\f{1}{2}\beta\sqrt{|3p-\rho |},
\end{align}
and
\begin{align}\label{f3}
(1+\alpha p)(2\dot{H}+3H^2)&=2\alpha H\dot{\rho }-\kappa p\nonumber\\&+\frac{1}{2}\alpha \left(\ddot{\rho }-\ddot{p}\right)-\f{2\beta p}{\sqrt{|3p-\rho |}},
\end{align}
respectively.

\subsubsection{High density regime}

In the high density cosmological regime the matter equation of state is given by the Zeldovich stiff causal equation of state, with $p=\rho$.  Then the field equations take the form
\be
3\left(1-\alpha \rho \right)H^2=\kappa \rho -\sqrt{2}\beta \sqrt{\rho },
\ee
and
\be
\left(1+\alpha \rho \right)\left(2\dot{H}+3H^2\right)=2\alpha H\dot{\rho }-\kappa \rho -\sqrt{2}\beta \sqrt{\rho },
\ee
respectively. For a small coupling $\alpha $, and by assuming $\alpha \rho \ll 1$, the field equations reduce to
\bea
3H^2&=&\kappa \rho -\sqrt{2}\beta \sqrt{\rho }, \label{T2}
 \\
\dot{H}&=&\alpha H\dot{\rho }-\kappa \rho, \label{T2b}
\eea
thus giving the evolution equation  for the density as
\be\label{T1}
\dot{\rho }(t)=\frac{4 \kappa  \rho ^{3/2}(t) \sqrt{3 \kappa  \rho (t)-3 \sqrt{2} \beta  \sqrt{\rho
   (t)}}}{\sqrt{2} \beta  \left[1-4 \alpha  \rho (t)\right]+2 \kappa  \sqrt{\rho (t)} \left[2 \alpha
   \rho (t)-1\right]}.
\ee
By neglecting the term $\alpha \rho $ compared to 1, and by series expanding the right hand side of Eq.~(\ref{T1}), to first order  we obtain
\be
\dot{\rho}=-2  \sqrt{3\kappa } \rho ^{3/2} \left(1-\frac{\beta }{\sqrt{2} \kappa
   \sqrt{\rho }}\right) \left(1+\frac{\beta }{2 \sqrt{2} \kappa  \sqrt{\rho
   }}\right),
\ee
with the general solution given by
\bea\label{rhoT}
\rho &=&\frac{\beta ^2 }{32 \kappa ^2} \Bigg\{3 \tanh \Bigg[\frac{3}{8} \Bigg(\frac{8}{3} \tanh
   ^{-1}\left(\frac{\beta \pm 4  \kappa  \sqrt{2\rho _0}}{3 \beta
   }\right)\nn
&&+\frac{\sqrt{6} \beta  t}{\sqrt{\kappa }}\Bigg)\Bigg]-1\Bigg\}^2,
\eea
where we have used the initial condition $\rho (0)=\rho _0$. After substituting the density given by Eq.~(\ref{rhoT}) into Eq.~(\ref{T2}), and performing a series expansion with respect to the time, to first order we obtain for the Hubble parameter
\bea
H(t)&=&\frac{\sqrt{\kappa  \rho _0-\sqrt{2} \beta  \sqrt{\rho
   _0}}}{\sqrt{3}} \nn
   &&+\frac{ \sqrt{2} \beta ^3-6 \sqrt{2} \beta  \kappa ^2
   \rho _0+8 \kappa ^3 \rho _0^{3/2}}{8 \kappa ^{3/2}
   \sqrt{\sqrt{\rho _0} \left(\kappa
   \sqrt{\rho _0}-\sqrt{2} \beta \right)}}t.
   \eea

In order to obtain a physical solution the parameters of the model must satisfy the constraint $\kappa
   \sqrt{\rho _0}>\sqrt{2} \beta$. For the scale factor of the Universe we obtain
   \bea
   a(t)&=&a_0\exp \Bigg\{ \frac{\sqrt{\kappa  \rho _0-\sqrt{2} \beta  \sqrt{\rho
   _0}}}{\sqrt{3}}t\nonumber\\
   &&+\frac{ \sqrt{2} \beta ^3-6 \sqrt{2} \beta  \kappa ^2
   \rho _0+8 \kappa ^3 \rho _0^{3/2}}{8 \kappa ^{3/2}
   \sqrt{\sqrt{\rho _0} \left(\kappa
   \sqrt{\rho _0}-\sqrt{2} \beta \right)}}\frac{t^2}{2}
   \Bigg\}.
   \eea
   In  the high density regime, and in the considered order of approximation, the expansion of the Universe is super-exponential, with the scale factor proportional to the exponential of $t^2$.

   \subsubsection{The pressureless matter fluid case}

   In the case of dust, having $p=0$, the gravitational field equations take the form
   \bea
   3\left(1-\alpha \rho \right)H^2&=&\kappa \rho +\frac{3}{2}\alpha H\dot{\rho }-\frac{\beta}{2}\sqrt{\rho },\label{101}  \\
   2\dot{H}+3H^2&=&2\alpha H\dot{\rho }+\frac{1}{2}\alpha \ddot{\rho },  \label{102}
   \eea
   respectively. We consider  the late time expansionary phase of the Universe, by assuming a de Sitter type form for the scale factor, $a(t)=\exp \left(H_0t\right)$, with $H_0={\rm constant}$.  Then Eq.~(\ref{102}) can be immediately integrated, to give
   \bea
  \rho (t)&=&e^{-4H_0\left(t-t_0\right)}\left(\frac{3}{8\alpha }-\frac{\rho _{01}}{4H_0}\right)+\frac{3H_0}{2\alpha }\left(t-t_0\right)  \nn
  &&+\rho _0+\frac{\rho _{01}}{4H_0}-\frac{3}{8\alpha },
   \eea
   where $\rho _0=\rho \left(t_0\right)$, and $\rho _{01}=\dot{\rho }\left(t_0\right)$. In the limit of large time the matter density is linearly increasing in time,  and hence this model does not have a physical late time de Sitter phase. Other types of solutions, including the matter dominated phase,  can be obtained through the detailed numerical study of the system of Eqs.~(\ref{101}) and (\ref{102}), which will not be performed here.

\subsection{Specific case III: $f=R\left(1+\alpha  R_{\mu\nu}T^{\mu\nu}\right)$}\label{C}

As a third example of a cosmological model  we consider the case in which the function $f$ is given by $f=R\left(1+\alpha  R_{\mu\nu}T^{\mu\nu}\right)$. The field equations  in this case are given by
\begin{align}
&\bigg[1+\alpha(R_{\alpha\beta}T^{\alpha\beta}-RL_m)\bigg]G_{\mu\nu}+\alpha\bigg[\Box(R_{\alpha\beta}T^{\alpha\beta})
     \nonumber\\
&+\nabla_\alpha\nabla_\beta(RT^{\alpha\beta})\bigg]g_{\mu\nu}-\alpha\nabla_\mu\nabla_\nu(R_{\alpha\beta}
T^{\alpha\beta})
   \nonumber\\
&+\f{1}{2}\alpha\Box(RT_{\mu\nu})+2\alpha RR_{\alpha(\mu}T^\alpha_{~\nu)}-\alpha\nabla_\alpha\nabla_{(\mu}\big[RT^\alpha_{~\nu)}\big]
   \nonumber\\
& -\bigg(\f{1}{2}\alpha R^2+8\pi G\bigg)T_{\mu\nu}-2\alpha RR^{\alpha\beta}\f{\partial^2 L_m}{\partial g^{\mu\nu}\partial g^{\alpha\beta}}=0.
\end{align}

 \subsubsection{The matter dominated phase}

As a first example of a cosmological solution of the field equations Eqs.~(\ref{co6}) and (\ref{co7}) we consider that the scale factor has a power law time evolution, $a=t^{\beta}$, where $\beta $ is a constant. In this case the field equations are
 \begin{eqnarray}\label{mod31}
27\,t \left( \beta-2/3 \right) \alpha\,{\beta}^{2}{\frac {d \rho \left( t \right)}{d
t}} -45\, \left( \beta-2/5 \right) t\alpha\,{
\beta}^{2}{\frac {d p \left( t \right)}{dt}}
    \nonumber  \\
+\left( 27\,\alpha\,{
\beta}^{2}-54\,\alpha\,{\beta}^{3}+\kappa\,{t}^{4}+27\,\alpha\,{\beta}
^{4} \right) \rho \left( t \right)
\nonumber\\
 -3\, \left( -9\,\alpha\, \left( {
\beta}^{2}-1+2\,\beta \right) p \left( t \right) +{t}^{2} \right) {
\beta}^{2}=0,
\end{eqnarray}
and
   \begin{widetext}
 \begin{align}\label{mod32}
\Bigg[ 9\,{t}^{2} &\left( \beta-2/3 \right) \alpha\,\beta\,{\frac {d^{2} \rho \left( t \right)}{d{t}^{2}}} -15\, \left( \beta-2/5 \right) {t}
^{2}\alpha\,\beta\,{\frac {d^{2} p \left( t \right) }{d{t}^{2}}} +30\,
 \left( \beta-4/5 \right) t\alpha\,\beta\, \left( \beta-1 \right) {
\frac {d \rho \left( t \right) }{dt}}
    \nonumber\\
&-18\, \left( {\beta}^{2}+4/3-11/3
\,\beta \right) t\alpha\,\beta\,{\frac {d p \left( t \right)}{dt}} +
 \left( -9\,\alpha\,{\beta}^{4}-111\,\alpha\,{\beta}^{2}+36\,\beta\,
\alpha-\kappa\,{t}^{4}+60\,\alpha\,{\beta}^{3} \right) p \left( t
 \right) \nonumber\\&-3\, \left( 3\, \left( -{\frac {29}{3}}\,\beta+4+{\beta}^{3}+
16/3\,{\beta}^{2} \right) \alpha\,\rho \left( t \right) +{t}^{2}
 \left( \beta-2/3 \right)  \right) \beta \Bigg] {t}^{-4}=0,
\end{align}
\end{widetext}
respectively. The general solution of these cosmological evolution equations involves an implicit differential equation for $p(t)$, obtained from Eq.~(\ref{mod32}).  Then $\rho(t)$ can be determined in terms of $p(t)$ from Eq.~(\ref{mod31}). In the particular case $\beta=2/3$, $p(t)$ is determined by the equation
\bea\label{mod33}
 && {\frac {d^{2} p \left( t \right) }{d{t}^{2}}} +{\frac {  648\alpha{
t}^{7}{\kappa}^{2}-672{\alpha}^{2}{t}^{3}\kappa   }{128{\alpha}^{3}-192{\alpha}^{2}
\kappa{t}^{4}-216\,\alpha\,{\kappa}^{2}{t}^{8}}}{\frac {d p \left( t \right)}{d
t}}
    \nonumber\\
&&-{\frac { 972\alpha {\kappa}^{2}{t}^{6}+81{\kappa}^{3}{t}^{10}-1536{
\alpha}^{2}\kappa{t}^{2}   }{128
{\alpha}^{3}-192{\alpha}^{2}\kappa {t}^{4}-216\alpha{\kappa}^{2
}{t}^{8}}}p \left( t \right)
  \nonumber\\
&&+{\frac {2\left(4\alpha -21\kappa {t}^{4} \right)}{16{\alpha}^{2}
-24\alpha \kappa {t}^{4}-27{\kappa}^{2}{t}^{8}}} =0 ,
\eea
and $\rho(t)$ can be determined from the equation
\begin{align}\label{}
\rho \left( t
 \right) ={\frac {16\alpha t\left[ {\frac {d p \left( t
 \right)}{dt}}  \right] -28\alpha\,p \left( t \right) +4{t}^{2}}{4
\alpha+3\kappa {t}^{4}}}.
\end{align}
In the limit of small $t$, $t\rightarrow 0$, Eq.~(\ref{mod33}) can be approximated as
\be
\frac {d^{2} p \left( t \right)}{d{t}^{2}}+\frac{1}{2\alpha }\approx 0,
\ee
giving for the time evolution of the pressure $p(t)=p_0+p_{01} \left(t-t_0\right)-\left(t-t_0\right)^2/4 \alpha $, where $p_0=p\left(t_0\right)$, and $p_{01}=\dot{p}\left(t_0\right)$. The energy density  for this decelerating, matter dominated phase, is given by
\be
\rho (t)\approx \frac{-28 \alpha p_0-6 t (2 \alpha p_{01}+t_{0})+7 t_{0} (4
   \alpha p_{01}+t_{0})+3 t^2}{4 \alpha +3 \kappa  t^4}.
    \ee

   \subsubsection{The de Sitter type phase of evolution}

   In the following, we investigate the cosmological solutions for the zero pressure matter filled Universe. The cosmological gravitational field equations are given by
\begin{align}\label{co6}
&-3 H^2+\kappa  \rho +\alpha  \bigg(18 H   \ddot{H}\rho+18 H \dot{H} \dot{\rho }\nonumber\\
&+54 H^2  \dot{H}\rho -9  \dot{H}^2\rho +27 H^3 \dot{\rho } +27H^4 \rho \bigg)=0,
   \end{align}
   and
   \bea\label{co7}
  && -2 \dot{H}-3 H^2+\alpha  \Bigg(6 \dddot{H} \rho +12 \ddot{H} \dot{\rho }+36 H  \ddot{H}\rho
   \nonumber\\
 &&  +6 \dot{H} \ddot{\rho } +  54 H \dot{H}
   \dot{\rho }+48 H^2  \dot{H}\rho +15  \dot{H}^2\rho +9 H^2 \ddot{\rho }
   \nonumber\\
   && +
   30 H^3 \dot{\rho }- 9 H^4 \rho \Bigg)=0,
   \eea
   respectively. The terms proportional to $\alpha $ in the generalized Friedmann equations (\ref{co6}) and (\ref{co7}) play the role of an effective supplementary density and pressure, which  may be responsible for the late time acceleration of the Universe.

Next, we look for a de Sitter type solution of Eqs.~(\ref{co6}) and (\ref{co7}), assuming that $H=H_0={\rm constant}$. Then the field equations take the form
   \bea
   -3H_0^2+\kappa \rho +27H_0^3\alpha \left(\dot{\rho }+H_0\rho \right)=0, \\
   -3H_0^2+H_0^2\alpha \left(9\ddot{\rho}+30H_0\dot{\rho}-9H_0^2\rho \right)=0,
   \eea
   respectively, leading to the following differential consistency condition for the matter density $\rho $,
   \be\label{eqf}
  9 \alpha  H_0^2 \ddot{\rho }+3 \alpha  H_0^3 \dot{\rho }-\left(36 \alpha  H_0^4  +\kappa \right) \rho =0.
   \ee

   The general solution of Eq.~(\ref{eqf}) is given by
   \bea
  \rho (t)&=& e^{-\frac{1}{6} H_0 \left(t-t_0\right)} \times \Bigg\{\frac{\sqrt{\alpha}  H_0 \left(H_0 \rho _0+6
   \rho _{01}\right) }{\sqrt{
   145 \alpha  H_0^4+4 \kappa }}\times \nonumber\\
  &&\times \sinh \left[\frac{ \sqrt{ 145
   \alpha  H_0^4+4 \kappa }}{6 \sqrt{\alpha } H_0}\left(t-t_0\right)\right]\nonumber\\
   && + \rho _0 \cosh
   \left[\frac{ \sqrt{145 \alpha  H_0^4+4 \kappa }}{6 \sqrt{\alpha
   } H_0}\left(t-t_0\right)\right]\Bigg\},
   \eea
where we have used the initial conditions $\rho \left(t_0\right)=\rho _0$ and $\dot{\rho }\left(t_0\right)=\rho _{01}$, respectively.
In order that the ordinary matter density decays exponentially for $t\geq t_0$, all the exponential terms must be negative, which imposes on $\alpha $ the constraint $\alpha <-\kappa /36H_0^4$. The high energy density regime of this model, corresponding to $p=\rho $, has similar properties with the $p=0$ case, that is, it admits a de Sitter phase, which can be obtained by analytical methods.

\subsection{$f\left(R,T,R_{\mu\nu}T^{\mu\nu}\right)$ gravity cosmological models with conserved energy-momentum tensor}\label{D}

We now consider cosmological models with a conserved energy-momentum tensor. For this case the relevant field equations are obtained in Section \ref{sec6} by using the Lagrange multiplier method, and are given by Eqs.~(\ref{lf}) and (\ref{encons}), respectively. For the isotropic and homogeneous FLRW Universe the energy conservation equation becomes
\be\label{co}
\dot{\rho}+3H(\rho+p)=0.
\ee
Assuming a barotropic equation  of state for the matter of the form $p(t)=\omega\rho(t)$, $\omega ={\rm constant}$,  and the ansatz $\lambda^\mu=\lambda(t)\delta^\mu_0$ for the Lagrange multiplier, the gravitational field equations with energy conservation are given by
\begin{widetext}
\be\label{l1}
3\bigg(f_R+\f{\omega-1}{2}\rho f_{RT}\bigg)\dot{H}+3\bigg(f_R+\f{3\omega^2+3\omega-2}{2}\rho f_{RT}\bigg)H^2+\bigg\{\bigg[(\omega+1)\lambda-\f{\omega-1}{2}\dot{f}_{RT}\bigg]\rho-\dot{f}_R\bigg\}H
+\rho\dot{\lambda}-\f{1}{2}f+8\pi G\rho=0,
\ee
and
\bea\label{l2}
&&\bigg(f_R+\f{3\omega^2-\omega-6}{2}\rho f_{RT}\bigg)\dot{H}+3\bigg(f_R-\f{3\omega^3+3\omega^2+2}{2}\rho f_{RT}\bigg)H^2+
\bigg\{\bigg[(3\omega^2-1)\dot{f}_{RT}+\f{\omega+3}{2}\lambda\bigg]\rho-2\dot{f}_R\bigg\}H
    \nonumber\\
&&-\ddot{f}_R-\f{1}{2}(\omega-1)\rho\ddot{f}_{RT}+\f{1}{2}(\omega+1)\rho\dot{\lambda}-(\omega+1)\rho f_T-\kappa \omega\rho-\f{1}{2}f=0,
\eea
\end{widetext}
respectively,  where we have eliminated $\dot{\rho }$ from the above equations by using the conservation equation Eq.~(\ref{co}).

As an example for cosmological applications we  consider the case where the function $f$ is given by
\be
f=R+\alpha R_{\mu\nu}T^{\mu\nu},
\ee
 where $\alpha ={\rm constant}$. In this case  Eqs.~(\ref{l1}) and (\ref{l2}) become
\be
3\bigg[\f{1}{2}(3\omega^2-1)\alpha\rho-1\bigg]H^2+3(\omega+1)\lambda \rho H+\rho\dot{\lambda}+\kappa \rho=0,
\ee
and
\bea
\bigg[2-\f{\alpha}{2}(3\omega^2-4\omega-3)\rho\bigg]\dot{H}+\f{9}{2}\alpha\omega(\omega+1)^2\rho H^2\nonumber\\
+\f{1}{2}(5\omega+3)\lambda\rho H+\f{1}{2}(1-\omega)\rho\dot{\lambda}+\kappa (\omega+1)\rho=0,
\eea
respectively.

\subsubsection{The high energy density phase}

In the high energy density limit we assume that the equation of state of the cosmological matter is the stiff causal equation of state, with $p=\rho$. Then the energy conservation equation gives
\be
p=\rho =\frac{\rho_0}{a^6}.
\ee

The field equations for the high density phase of the evolution of the Universe are given by
\be\label{118}
3\left(\alpha \rho -1\right)H^2+6\lambda H\rho +\rho \dot{\lambda }+\kappa \rho =0,
\ee
and
\be\label{119}
(1+\alpha \rho )\dot{H}+9\alpha \rho H^2+2\lambda \rho H+\kappa \rho =0,
\ee
respectively. By assuming that $\alpha \rho  \gg 1$, Eqs.~(\ref{118}) and (\ref{119}) become
\bea
3\alpha H^2+6\lambda H + \dot{\lambda }+\kappa &=&0, \label{120} \\
\alpha \dot{H}+9\alpha  H^2+2\lambda  H+\kappa  &=&0.  \label{120b}
\eea
For $H=H_0={\rm constant}$, and $\lambda =\lambda _0={\rm constant}$, and for $\alpha <0$, Eqs.~(\ref{120})-(\ref{120b}) have the solution
\be\label{121}
H_0=\frac{1}{2}\sqrt{\frac{\kappa }{3|\alpha |}},\lambda _0=\frac{1}{4}\sqrt{3\kappa |\alpha |}.
\ee
Therefore in the  $f\left(R,T,R_{\mu\nu}T^{\mu\nu}\right)$ gravity with energy conservation a de Sitter type phase does exist during the high density regime of the cosmological evolution of the Universe.
From the above equations we obtain the relation between the Lagrange multiplier and the Hubble parameter as
\be
\lambda _0=\frac{1}{8}\frac{\kappa }{H_0}.
\ee

\subsubsection{The pressureless matter case}

In the case of dust matter, i.e., $\omega=0$, from the conservation of the energy-momentum tensor we obtain the density of the Universe as
\be
\rho =\frac{\rho _0}{a^3}.
\ee
The gravitational field equations with the conservation of energy-momentum  and dust matter take the form
\be\label{cc1}
-3\left(1+\frac{\alpha }{2}\rho \right)H^2+3\lambda \rho H+\rho \dot{\lambda }+\kappa \rho =0,
\ee
and
\be\label{cc2}
\left(2+\frac{3\alpha }{2}\rho \right)\dot{H}+\frac{3}{2}\lambda \rho H+\frac{1}{2}\rho \dot{\lambda}+\kappa \rho =0,
\ee
respectively. From Eqs.~(\ref{cc1}) and (\ref{cc2}) we immediately obtain
\be\label{cc3}
2\left(2+\frac{3\alpha }{2}\rho \right)\dot{H}+3\left(1+\frac{\alpha }{2}\rho \right)H^2+\kappa \rho =0.
\ee

In the limit of large densities $\alpha \rho \gg 1$, Eq.~(\ref{cc3}) becomes
\be
 3\alpha \dot{H}+3\frac{\alpha }{2}H^2+\kappa =0,
 \ee
 with the general solution given for $\alpha >0$ by
 \be
H(t)=\sqrt{\frac{2\kappa }{3\alpha }}  \tan \left[\tan ^{-1}\left(\sqrt{\frac{3\alpha }{2\kappa }} H_0 \right)+\sqrt{\frac{\kappa
   }{6\alpha }} \left(t_0-t\right)\right],
   \ee
 where we have used the initial condition $H\left(t_0\right)=H_0$. For $\alpha <0$ we obtain
 \be
 H(t)=\sqrt{\frac{2\kappa }{3\alpha }}  \tanh \left[\tanh ^{-1}\left(\sqrt{\frac{3\alpha }{2\kappa }} H_0 \right)+\sqrt{\frac{\kappa
   }{6\alpha }} \left(t-t_0\right)\right].
   \ee

 For $\alpha >0$, the time evolution of the scale factor is given by
 \be
 a(t)=a_0\cos ^2\left[\sqrt{\frac{\kappa }{6\alpha }} \left(t-t_0\right)-a_{>}\right],
 \ee
 where $a_0$ is an arbitrary constant of integration and we define
 $$
 a_{>}\equiv\tan ^{-1}\left(\sqrt{\frac{3\alpha }{2\kappa }} H_0 \right).
 $$
 One can see that in this case we have a bouncing universe. The deceleration parameter can be obtained as
$$
q=\left( \sec\left[\sqrt{\frac{2\kappa }{3\alpha }} \left(t-t_0\right)-2a_{>}\right] -1\right)^{-1},
$$
which is negative for
\begin{align}
T<t<T+\sqrt{\f{3\alpha}{2\kappa}}\pi,
\end{align}
where we have defined
$$
T=\sqrt{\f{3\alpha}{2\kappa}}\left(\f{2n+1}{2}\pi+2a_{<}+\sqrt{\f{2\kappa}{3\alpha}}t_0 \right),
$$
and $n=0,1,2,...$.
For $\alpha <0$ the scale factor takes the form
 \be
 a(t)=a_0\cosh ^2\left[\sqrt{\frac{\kappa }{6\alpha }} \left(t-t_0\right)+a_{<}\right],
 \ee
where in this case
$$
a_{<}\equiv\tanh ^{-1}\left(\sqrt{\frac{3\alpha }{2\kappa }} H_0 \right).
$$
The deceleration parameter can then be obtained as
$$
q=\left(1- \textmd{sech}\left[\sqrt{\frac{2\kappa }{3\alpha }} \left(t-t_0\right)+2a_{<}\right] \right)^{-1},
$$
In this case the deceleration parameter is always positive and we have a decelerating universe.

 In the opposite limit of small densities $\alpha \rho \ll 1$, Eq.~(\ref{cc3}) takes the form
 \be
 4\dot{H}+3H^2+\frac{\kappa \rho _0}{a^3}=0,
 \ee
 with the general solution given by
 \bea
 a(t)&=&\frac{\left\{\left(3 a_0^3 H_0^2-\kappa \rho _0\right) \right\}^{2/3}}{4 a_0 \left(6 a_0^3 H_0^2-2 \kappa  \rho _0\right)^{2/3}}\times \nonumber\\
 && \hspace{-1.25cm} \times \left[a_0^3 (3 H_0 \left(t-t_0\right)+4)^2-3 \kappa  \rho _0
   \left(t-t_0\right)^2\right]^{2/3},
   \eea
where we have used the initial conditions $a\left(t_0\right)=a_0$, and $\dot{a}\left(t_0\right)=a_0H_0$, where $H_0=H\left(t_0\right)$. The deceleration parameter can be obtain as
\begin{align}
q=-\f{1}{4}+\f{4\kappa\rho_0a_0^3}{\big[\kappa\rho(t-t_0)-a_0^3H_0(4+3(t-t_0)H_0)\big]^2}
\end{align}
In this case one can easily see that for
\begin{align}
0<t<4\f{a_0^3H_0\mp\sqrt{\kappa\rho_0a_0^3}}{\kappa\rho_0-3a_0^3H_0^2}+t_0
\end{align}
or
\begin{align}
t>4\f{a_0^3H_0\pm\sqrt{\kappa\rho_0a_0^3}}{\kappa\rho_0-3a_0^3H_0^2}+t_0
\end{align}
we have an accelerating universe. We note that the upper and lower signs refer to the cases $\kappa>3a_0^3H_0^2/\rho_0$ and $0<\kappa<3a_0^3H_0^2/\rho_0$ respectively.

The general solution of the cosmological field equations (\ref{cc1}) and (\ref{cc2}) can be obtained as
\be
t=\f{1}{\sqrt{16\pi G\rho_0}}\int\f{Aa^{1/2}\textmd{d}a}{\sqrt{a_0-B}},\qquad B=\int A^{-10}\textmd{d}a
\ee
where $a_0$ is an integration constant and we have denoted
 \be
 A=(3\alpha\rho_0+4a^3)^{1/12}.
 \ee
  One can then obtain the Lagrange multiplier $\lambda$ in terms of $a$ as
\be\label{138}
\lambda=\f{1}{2\rho_0 a^3}\int a^3\bigg(3\alpha\rho_0 H^2+6H^2 a^2-2\kappa \rho_0\bigg)\textmd{d}t.
\ee

In classical mechanics, the Lagrange multiplier has the meaning of the force that keeps the constraint on the mechanical system. In the models with energy conservation,  we would like to conserve the energy of the ordinary matter, which amounts to provide some energy to the gravitational system. Eq.~(\ref{138}) can be written in a differential form as
\be
\frac{1}{a^3}\frac{d}{dt}\left(\lambda a^3\right)=\frac{1}{2\rho _0}H^2\bigg(3\alpha\rho_0 +6 a^2-2\frac{\kappa \rho_0}{H^2}\bigg),
\ee
showing that the time variation  of the Lagrange multiplier density is proportional to the square of the Hubble parameter.

 \section{Discussions and final remarks}\label{sec8}

In this paper we have extended the work initiated in \cite{Pop} and \cite{fRT} by considering a more general gravitational action in which the Lagrangian of the field explicitly depends not only on $R$ and $T$, but also on the contraction of the matter energy-momentum tensor with the Ricci tensor. The gravitational field equations have been obtained in the metric formalism in two cases, corresponding to a non-conservative and conservative physical system, respectively. In order to impose the condition of the conservation of the energy-momentum tensor we have used a Lagrange multiplier method, which implies the introduction of a new vector field  in the gravitational action. The equation of motion of massive test particles was derived in the non-conservative case and so was its Newtonian limit, corresponding to weak gravitational fields and low velocities.  A density-dependent supplementary acceleration, acting on massive test particles, is induced in the presence of a non-minimal
coupling between geometry and matter. The extra force on massive particles generated by the geometry-matter  coupling is always present, even in the case $L_m=p$, and causes a deviation from geodesic paths. The presence of the extra force could
explain the properties of the galactic rotation curves without resorting to the dark matter hypothesis.
It is interesting to note that this supplementary acceleration is also proportional to the matter density gradient, tending to zero for constant density  self-gravitating systems. A similar dependence on the gradient of the Newtonian gravitational potential also appears in the generalized Poisson equation.

The viability of the theory was studied  by examining  the stability of the theory with respect to local perturbations. In pure $f(R)$ gravity, a fatal instability develops on time scales of the order of $10^{−26}$ s when the function $f(R)$ satisfies the condition
$f''(R) < 0$. This instability, called the “Dolgov-Kawasaki” instability, was discovered in the prototype model $f(R) = R - \mu ^4/R$, with $\mu \sim H_0\sim 10^{-33}$ eV \cite{DK},  which is therefore ruled out. In the present case,  the condition of the stability with respect to the local perturbations can be formulated as $f_{RR}\left(R_0\right)- \left(\rho _0-T_0/2\right)f_{RT,R}\left(R_0\right)\geq 0$, where $R_0$ is the background Ricci scalar.

The cosmological implications of the  theory were also  investigated for both conservative and non-conservative theories. For this study we have adopted four functional forms for $f\left(R,T,R_{\mu\nu}T^{\mu\nu}\right)$. In the non-conservative case we have shown  that for two choices of the function $f$, the gravitational field equations admit an exponential, de Sitter type solution. Therefore matter-geometry coupling may be responsible for the late time acceleration of the Universe, as suggested by the observation of the high redshift supernovae \cite{Riess}. An interesting solution of the field equations was  obtained in the case of a conservative model with $f\left(R,T,R_{\mu\nu}T^{\mu\nu}\right)=R+\alpha R_{\mu \nu }T^{\mu \nu }$. In
this case if the coupling constant $\alpha >0$, the solution has an oscillatory behavior, with alternating expanding and collapsing phases. For $\alpha <0$, the scale factor of the Universe has a hyperbolic cosine type dependence. We have also investigated models containing the square root of the trace of the energy-momentum tensor. In this case in the high density limit the Universe has a super-accelerated expansion, but no de Sitter type phase can be obtained analytically.

Work along similar lines has been done independently in \cite{Odintsovetal}, although in a different setting, and with a different focus, with mainly the cosmological aspects of the theory being investigated. Indeed, the authors of \cite{Odintsovetal} mainly considered the accelerating solutions of the $f(R,T,R_{\mu\nu}T^{\mu\nu})$ theory, and attempted to find the functional form of $f$ analytically. On the other hand we dealt with the other aspects of the theory, including  the motion of a test body in the gravitational field, as well as the Newtonian limit and the generalized Poisson equation. We have also considered some cosmological solutions for the model. An important new result in our work is the use of the Lagrange multiplier method to implement energy-momentum conservation. In both our work and in \cite{Odintsovetal}, the Dolgov-Kawasaki instability was explored, and the same results were obtained.

The field equations of $f\left(R,T,R_{\mu\nu}T^{\mu\nu}\right)$ gravity are extremely complex.  For different choices of the function $f$, cosmological solutions with many types of qualitative behaviors can be obtained. These models can be used to explain the late acceleration of the Universe, without resorting to the cosmological constant, or to the dark energy. On the other hand, this theory can open a new perspective on the very early stages of the evolution of the Universe, and may provide an alternative to the inflationary paradigm, which is facing very serious challenges due to the recently released Planck results.

\section*{Acknowledgments}
We thank S. D. Odintsov for helpful comments and suggestions on an earlier version of the manuscript.
FSNL acknowledges financial support of the Funda\c{c}\~{a}o para a Ci\^{e}ncia e Tecnologia through the
grants CERN/FP/123615/2011 and CERN/FP/123618/2011. Z. Haghani, H. R. Sepangi and S. Shahidi would like to thank the Research Council of Shahid Beheshti University for financial support.

\end{document}